%% file: growth_511keV_nat.tex
\documentclass[openany]{nature}
\usepackage{graphicx} 
\usepackage{geometry}
\usepackage{multirow}
\usepackage{url}

\usepackage{hyperref}

\title{Photonuclear Reactions in Lightning Discovered from Detection of Positrons and Neutrons}
\author{
Teruaki Enoto$^{1*}$,
Yuuki Wada$^{2,3}$,
Yoshihiro Furuta$^2$,
Kazuhiro Nakazawa$^{2,4}$,
Takayuki Yuasa$^5$,
Kazufumi Okuda$^2$,
Kazuo Makishima$^6$,
Mitsuteru Sato$^7$,
Yousuke Sato$^8$,
Toshio Nakano$^3$,
Daigo Umemoto$^9$
\& Harufumi Tsuchiya$^{10}$
}
\begin{document}

{\bf Comment:} This manuscript is an initial-submission version of ``Photonuclear Reactions Triggered by Lightning Discharge" which was accepted for publication in the 23 November 2017 issue of Nature as a Letter (DOI 10.1038/nature24630). This manuscript has not undergone the peer review process. See the accepted version at the Nature website: \href{http://dx.doi.org/10.1038/nature24630}{http://dx.doi.org/10.1038/nature24630}

\clearpage

\maketitle
\input{original_commands.tex}
\begin{affiliations}
    \item The Hakubi Center for Advanced Research and Department of Astronomy, Kyoto University, Kyoto 606-8302, Japan
    \item Department of Physics, Graduate School of Science, The University of Tokyo, Tokyo 113-0033, Japan
    \item High Energy Astrophysics Laboratory, RIKEN Nishina Center, Saitama 351-0198, Japan
    \item Research Center for the Early Universe, The University of Tokyo, Tokyo 113-0033, Japan
    \item 55 Devonshire Road 239855, Singapore
    \item MAXI Team, The Institute of Physics and Chemical Research (RIKEN), Saitama 351-0198, Japan
    \item Graduate School of Science, Hokkaido University, Sapporo 060-0808, Japan
    \item Department of Applied Energy, Graduate School of Engineering, Nagoya University, Aichi 464-8603, Japan
    \item Advanced Institute for Computational Science, RIKEN, Hyogo 650-0047, Japan
    \item Nuclear Science and Engineering Center, Japan Atomic Energy Agency, Ibaraki 319-1195, Japan
\end{affiliations}
\begin{abstract}
Lightning and thundercloud are the most dramatic
    natural particle accelerators on the Earth~\cite{2012SSRv..173..133D}.
Relativistic electrons accelerated by electric fields therein
	emit bremsstrahlung gamma rays~\cite{GUREVICH1992463, 2012JGRA..117.2308D},
	which have been detected
	at ground observations~\cite{GRL:GRL14745, 2002JGRD..107.4324T, 2004GeoRL..31.5119D,  2007PhRvL..99p5002T, PhysRevD.82.043009, 2012JGRA..11710303D},
	by airborne detectors~\cite{GRL:GRL2924, JGRD:JGRD4405, 2015JPlPh..81d4705D},
    and as terrestrial gamma-ray flashes (TGFs)
    from space~\cite{1994Sci...264.1313F, 2005Sci...307.1085S, PhysRevLett.106.018501}.
The energy of the gamma rays is sufficiently high
    to potentially invoke atmospheric photonuclear reactions
    $^{14}$N~($\gamma$,~n)~$^{13}$N~\cite{2007Ge&Ae..47..664B, 2010JGRA..115.0E19C, 2014PhRvD..89i3010B, 2015JPlPh..81d4705D, JGRA:JGRA20239},
    which would produce neutrons
    and eventually positrons via beta-plus decay of generated unstable radioactive isotopes,
    especially $^{13}$N.
However, no clear observational evidence for the reaction
    has been reported to date~\cite{JGR:JGR15185, 1985Natur.313..773S, JGRA:JGRA14836}.
Here we report the first detection of neutron and positron signals
    from lightning with a ground observation.
During a thunderstorm on 6 February 2017 in Japan,
    a TGF-like intense flash ($<$~1~ms) was detected at our monitoring sites
    0.5--1.7~km away from the lightning.
The subsequent initial burst quickly subsided
    with an exponential decay constant of 40--60~ms,
    followed by a prolonged line emission
    at $\sim$~0.511 megaelectronvolt (MeV), lasting for a minute.
The observed decay timescale and spectral cutoff at $\sim$~10~MeV of the initial emission
    are well explained with de-excitation gamma rays
    from the nuclei excited by neutron capture.
The centre energy of the prolonged line emission
    corresponds to the electron-positron annihilation,
    and hence is the conclusive indication of positrons produced after the lightning.
Our detection of neutrons and positrons is unequivocal evidence
    that natural lightning triggers photonuclear reactions.
No other natural event on the Earth is known to trigger photonuclear reactions.
This discovery places lightning as only the second known natural channel on the Earth
    after the atmospheric cosmic-ray interaction,
    in which isotopes, such as $^{13}$C, $^{14}$C, and $^{15}$N, are produced.

\end{abstract}

\begin{figure}
\begin{center}
\includegraphics[width=0.5\hsize,bb=3 -10 486 539]{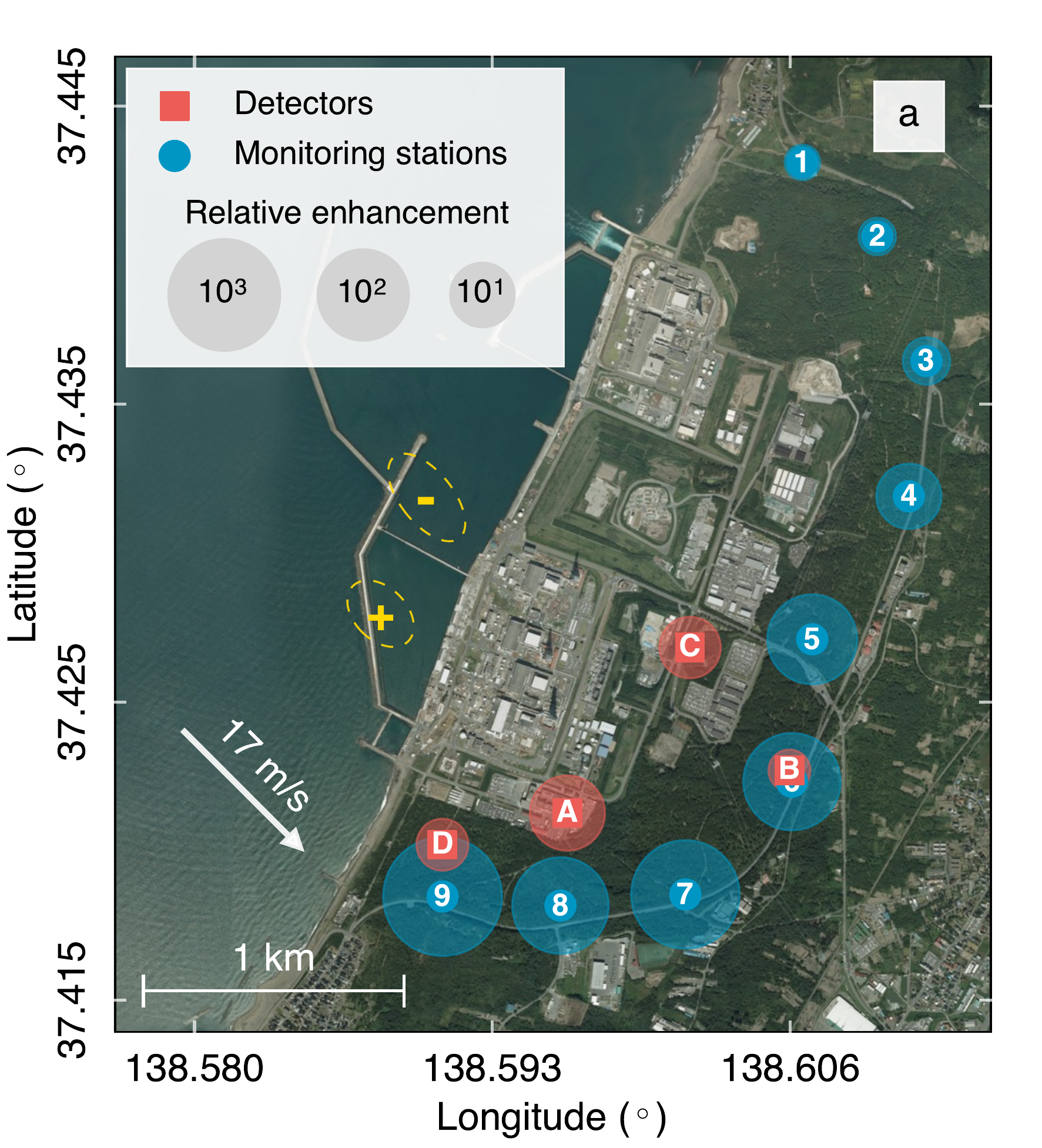}  
\includegraphics[width=0.6\hsize,bb=5 6 560 293]{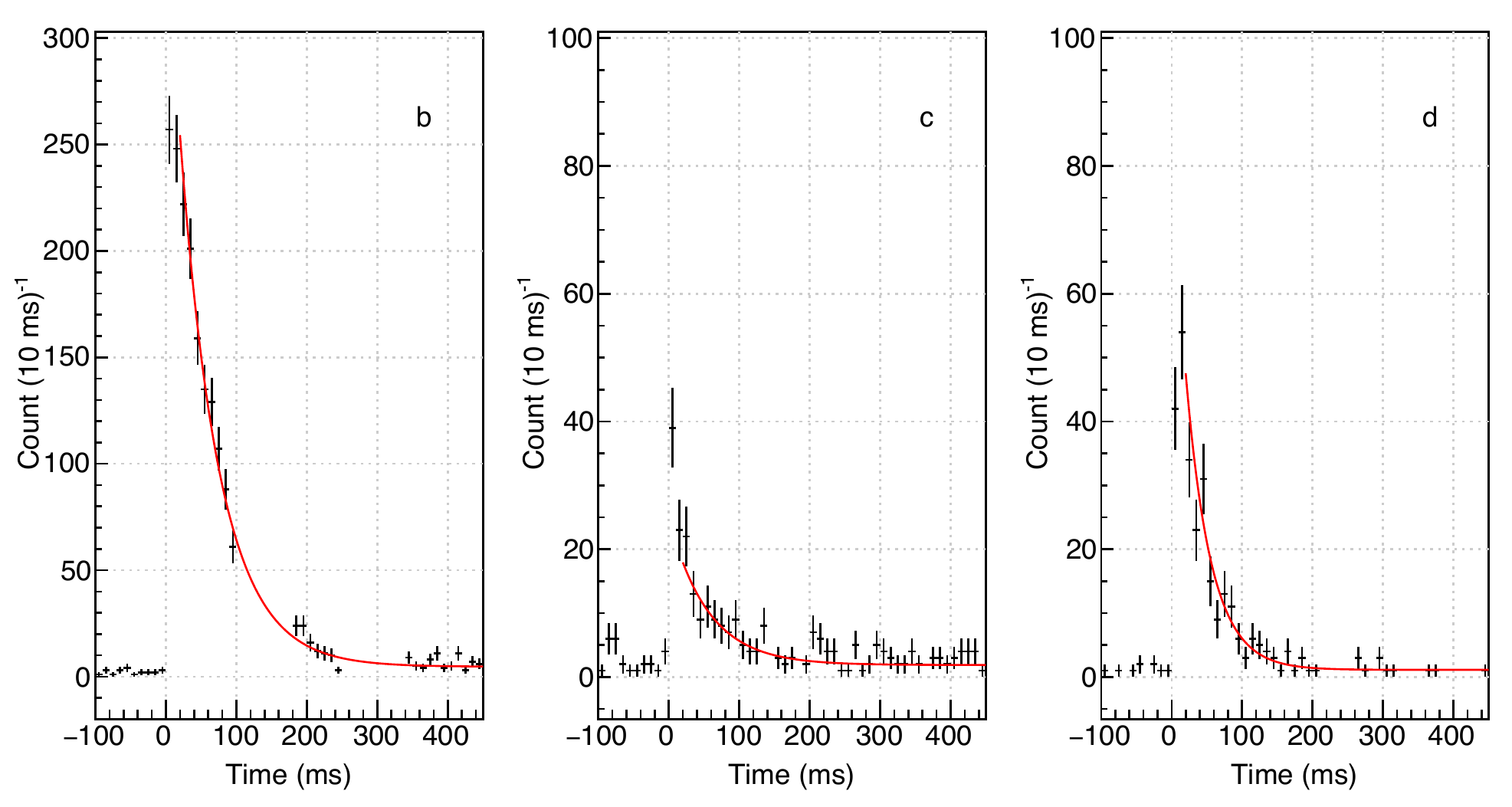}  
\caption{\textbf{Lightning discharges and associated sub-second decaying high-energy radiation.}
\textbf{a},
Aerial photograph of our observation site, Kashiwazaki-Kariwa, Niigata.
The two yellow dashed-lines show the positional error circles
    of the negative (``$-$'' symbol) and positive (``$+$'')
    cloud-to-ground discharges.
The red square and blue circles
    show the locations of our radiation detectors
    and radiation-monitoring stations, respectively.
The radius and the values in the legend of the overlaid circle on each point
    represents the intensity of the radiation enhancement at the lightning,
    relative to the environmental backgrounds
    averaged over $\sim$~10~min before and after the event.
The arrow shows the local wind direction at the time of the lightning.
\textbf{b--d},
10-ms-binned count-rate histories (black crosses for $\pm 1 \sigma$ errors)
    before and after the lightning (the time origin of the panels),
    with the detectors A (panel \textbf{b}, $>$~0.35~MeV),
    B (\textbf{c}, $>$~0.35~MeV),
    and C (\textbf{d}, $>$~1.2~MeV).
Red lines show the best-fitting model functions of an exponential decay.
The data loss due to dead time is corrected in the detector A,
    whereas the baseline undershoot is corrected for none of the detectors
    (see Methods ``Initial flash'').
The data gap for the detector A is due to overflow of memory buffer in the ADC board.
The detector D is not used here due to the undershoot (see Methods ``Initial flash").
\label{fig:fig1}    
}
\end{center}
\end{figure}

Winter thunderstorms along the coast of the Sea of Japan
    are much feared natural phenomena for the locals~\cite{rakov2003lightning}.
For the same reasons, the area in winter season is ideal
    for observing high-energy phenomena from lightning and thunderclouds \textit{i.e.},
    (1) powerful thunderstorms are frequent,
    (2) cold temperature lowers the cloud-base altitude
    down to 0.2--0.8~km~\cite{GotoNarita1992, rakov2003lightning},
    which allows gamma rays from it, if emitted, to reach the ground easily.
With the aim of detecting gamma rays from winter thunderclouds,
    we started the ``GROWTH project'' in 2006~\cite{2007PhRvL..99p5002T, 2011JGRD..116.9113T, Umemoto2016}
    (see Methods ``GROWTH collaboration''),
    which was later partly funded by crowdfunding.
During the winter of 2016--2017,
    we operated newly installed four radiation detectors A--D
    at the Kashiwazaki-Kariwa nuclear power station in Niigata (Fig~\ref{fig:fig1}a).

On 6 February 2017,
    a pair of negative and positive cloud-to-ground discharges
    occurred at 08:34:06.0027 UTC
    under a winter thunderstorm (see Methods ``Lightning discharges'').
All the four detectors, located 0.5--1.7~km away from the place of the discharges,
    simultaneously recorded an intense radiation
    lasting for $\sim$~200~ms (Fig~\ref{fig:fig1}). 
The radiation monitoring stations operated by the power plant
    also recorded this flash (see Fig~\ref{fig:fig1}a and Methods ``Radiation monitors'').
The analogue outputs of phototube amplifier exhibited strong undershoot (or equivalent)
    at the beginning of the event
    for roughly 40, 20, 20, and 300~ms
    in the detectors A--D, respectively (see Methods ``Initial flash'').
These undershoots are an instrumental response
    to intensive light outputs from scintillation crystals
    that largely exceeded the nominal instrumental dynamic range,
    which was triggered by a very short ($<$~1~ms) and strong gamma-ray flash,
    resembling a downward TGF~\cite{JGRD:JGRD53038}.
In the following analyses,
    we define $t$ as the elapsed time
    from the epoch of the rise of the initial radiation flash.

\begin{figure}
\begin{center}
\includegraphics[width=0.6\hsize,bb=4 1 317 294]{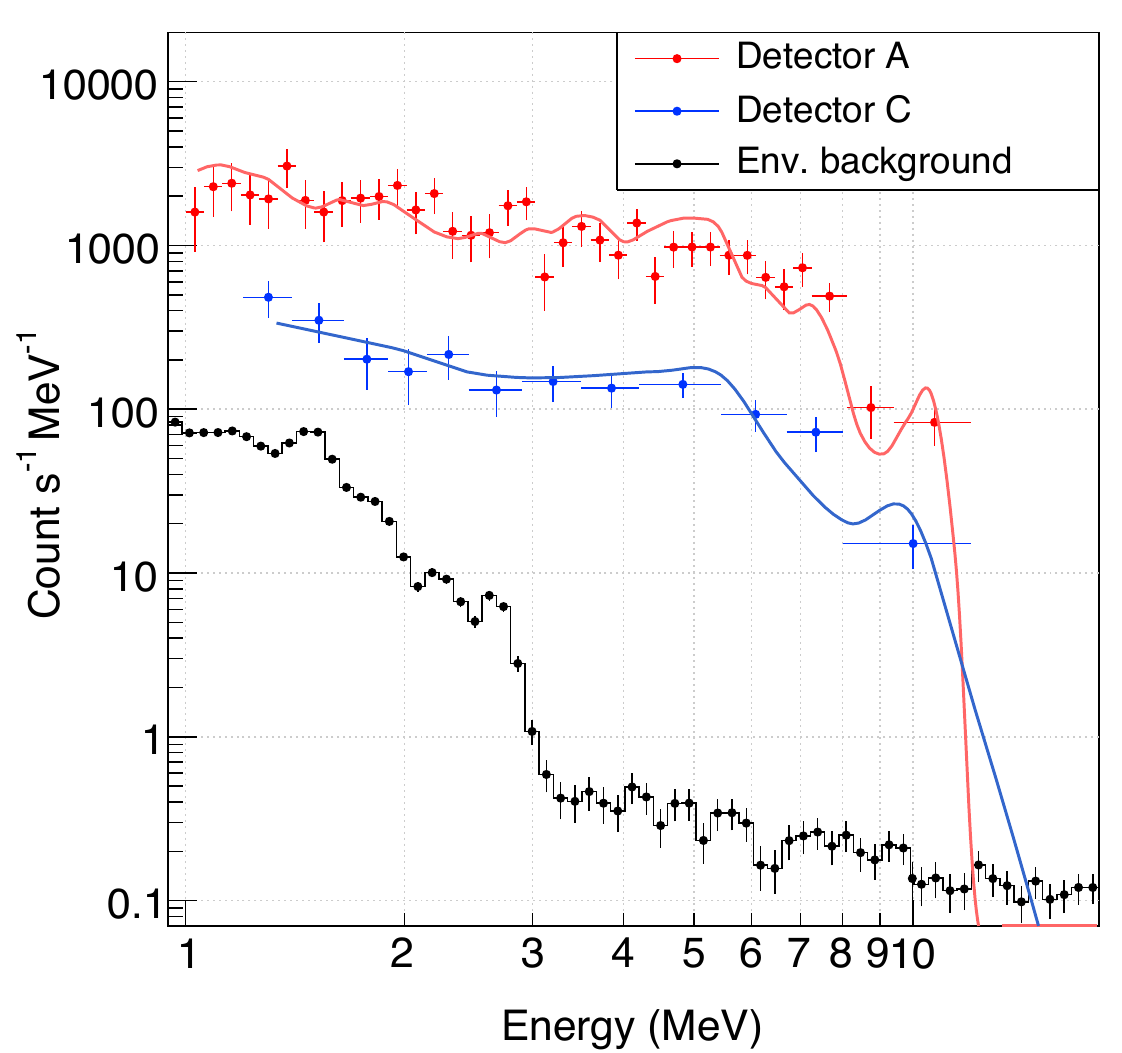}
\caption{\textbf{De-excitation gamma-ray spectra of the sub-second decaying radiation.}
Background-subtracted radiation spectra
    of the detectors (red crosses) A
    and (blue) C,
    compared with the simulated de-excitation gamma-ray spectra (solid lines). 
The events are accumulated over 40~$< t <$~100~ms and 20~$< t <$~200~ms, respectively.
The background spectrum is also plotted for comparison,
    extracted from $-10 < t < -130$~s and 90~$< t <$~210~s.
The read-out dead time is corrected for the detector A,
    whereas the instrumental response is inclusive.
The error bars are in $\pm 1 \sigma$.
\label{fig:fig2}
}
\end{center}
\end{figure}

After the initial surge of signals
    and after the nominal operation status of the amplifier
    had been restored (see Methods ``Initial flash"),
    all the detectors recorded a sub-second decaying radiation (Fig.~\ref{fig:fig1}b--d).
The time profile is fitted satisfactorily by an exponential form
    with a decay constant of 40--60~ms.
The recorded event rates with the detectors A--C
    above 3~MeV during the sub-second radiation
    are 2--3 orders of magnitude higher than the environmental background.
The spectra show a power-law shape with a photon index $\Gamma \sim$~0.5
    below a sharp cutoff at 7--10~MeV (Fig.~\ref{fig:fig2}).
These time profile and spectral shape suggest that
    the origin is different from the bremsstrahlung gamma rays
    in thundercloud electric fields 
    reported in the past~\cite{2007PhRvL..99p5002T, 2011JGRD..116.9113T},
    which showed Gaussian-like time profiles lasting for a minute
    and had an energy spectrum with a steeper slope $\Gamma \sim$~1--2
    and less sharp cutoff at around 20~MeV.

\begin{figure}
\begin{center}
\includegraphics[width=0.7\hsize,bb=6 3 314 294]{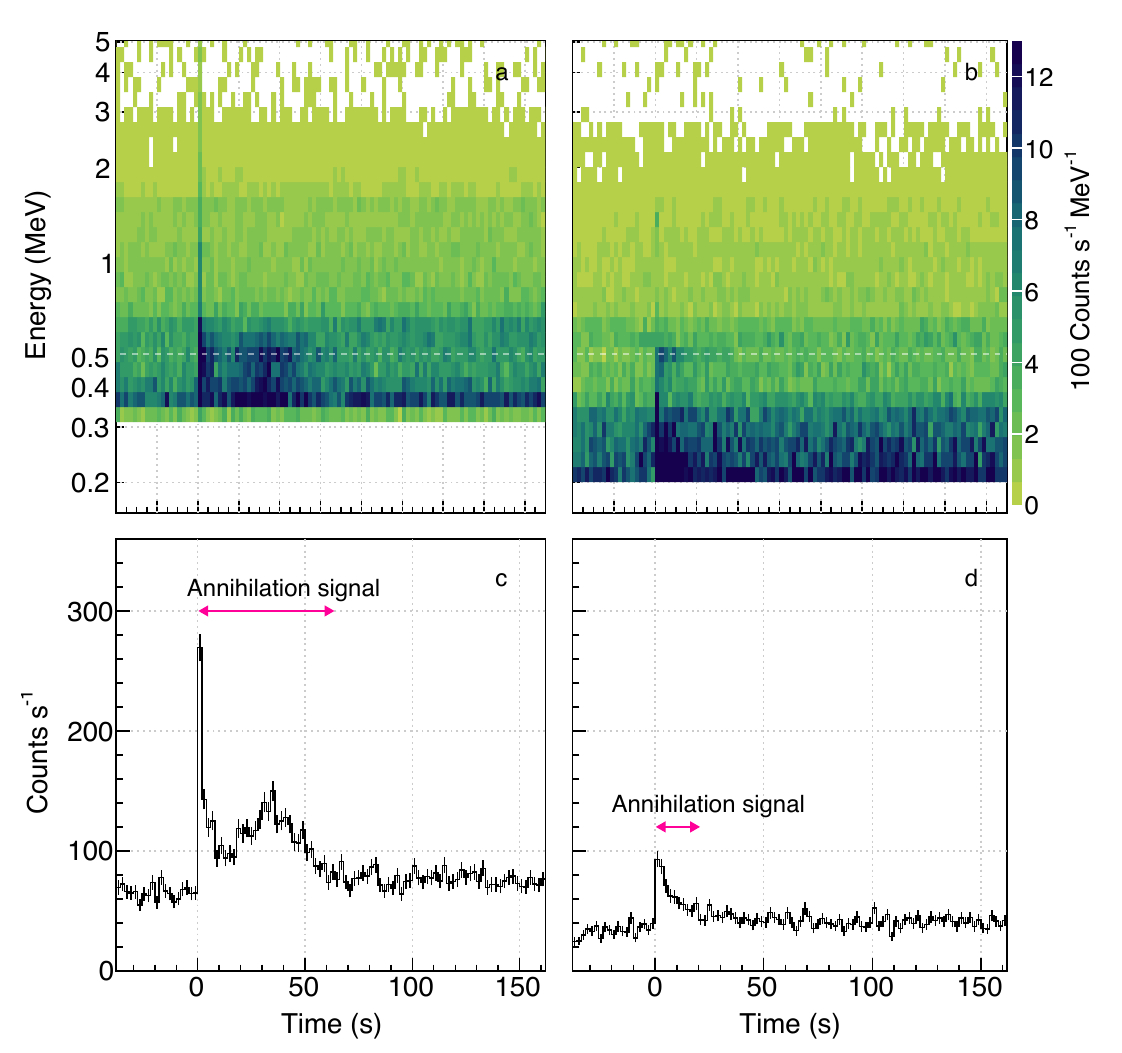}  
\caption{\textbf{Count-rate histories during the annihilation signal.}
\textbf{a, b},
The gradient coloured maps show
    the detected counts in a unit time and energy (counts~s$^{-1}$~MeV$^{-1}$)
    on the two-dimensional histograms of time (abscissa, 1-s binning)
    and energy (ordinate)
    for the detectors A (panel \textbf{a}) and D (\textbf{b}).
The horizontal white dashed-line indicates 0.511~MeV.
The dominant backgrounds above and below 3~MeV
    are cosmic-ray induced components
    and radiation from environmental radioactive nuclei, respectively.
\textbf{c, d},
2-s-binned count-rate histories of the 0.35--0.60~MeV band
    of the detectors A (panel \textbf{c}) and D (\textbf{d}).
Pink arrows in the panels \textbf{c} and \textbf{d} indicate
    the time durations of 1.0~$< t <$~63~s and 1.0~$ < t <$~20~s, respectively,
    from which the annihilation signals were detected
    and the spectra (Fig.~\ref{fig:fig4}) are accumulated.
The error bars are in $\pm 1 \sigma$.
\label{fig:fig3}
}
\end{center}
\end{figure}

\begin{figure}
\begin{center}
\includegraphics[width=0.6\hsize,bb=7 5 302 297]{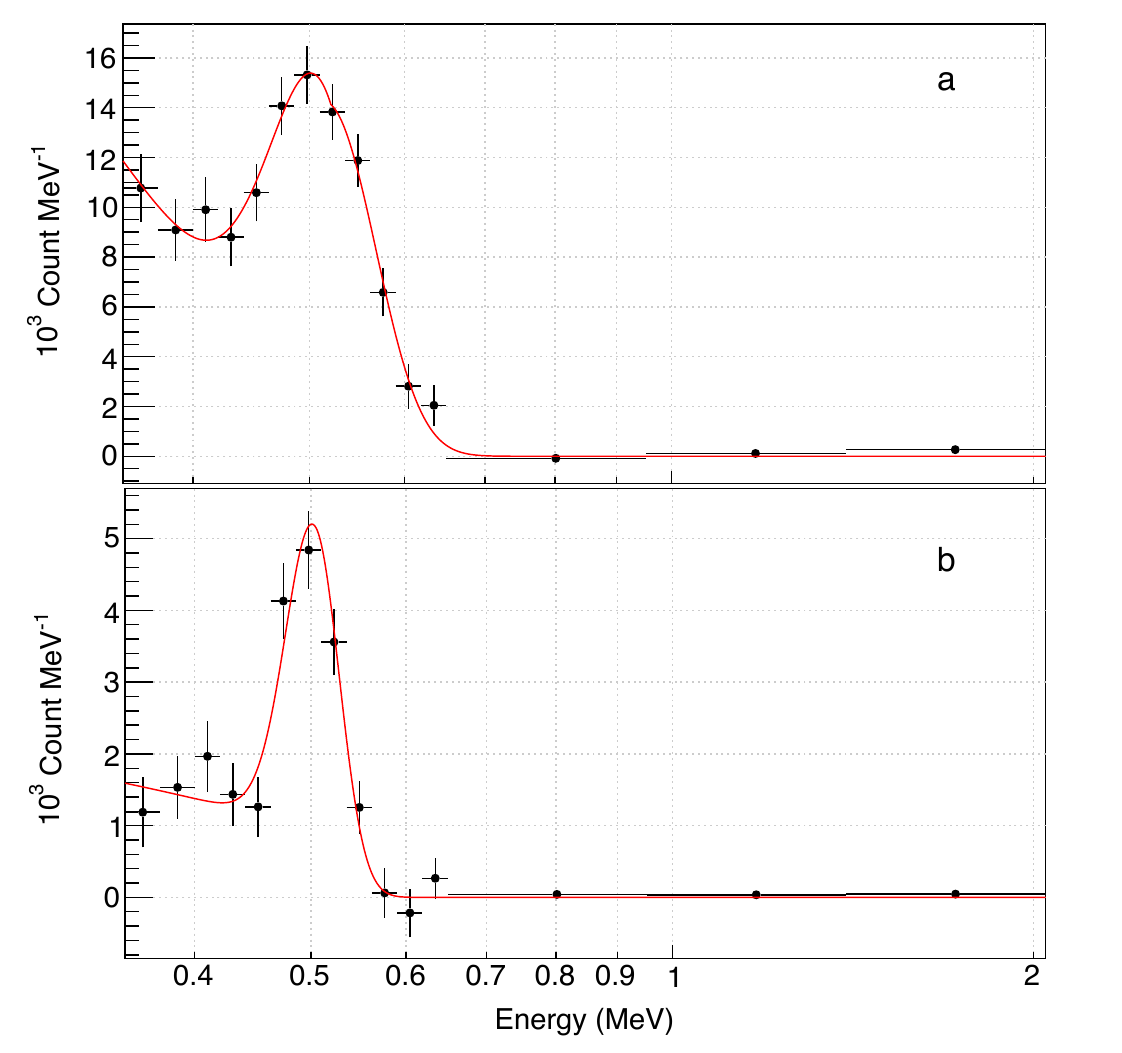}  
\caption{\textbf{Gamma-ray spectra during the prolonged annihilation signal.}
Panels \textbf{a} and \textbf{b} are for the detectors A and D, respectively.
The background is subtracted from the same time span of Fig.~\ref{fig:fig2}.
The detector response is inclusive.
The error bars are in $\pm 1 \sigma$.
See the panels \textbf{c} and \textbf{d} in Fig.~\ref{fig:fig3}
    for the event-accumulating time regions
    for the detectors A and D, respectively.
Red lines show the best-fitting empirical models
    consisting of a Gaussian line profile plus a power-law continuum,
    the latter of which represents the Compton scattering component from the former
    (see also Methods ``Positrons and annihilation" for a physically-based model).
\label{fig:fig4}
}
\end{center}
\end{figure}

Following the sub-second radiation,
    the 0.35--0.60~MeV count rates with the detectors A and D (Fig.~\ref{fig:fig3}a, b)
    increased for up to a minute.
Fig.~\ref{fig:fig4} shows the energy spectra during the period of the enhancement.
The most striking feature in the spectra
    is a prominent emission line at $\sim$~0.51~MeV,
    which is in very good agreement with the energy of 0.511~MeV
    for electron-positron annihilation.
The centre energies of the Gaussian line profiles were determined to be
    0.515~$\pm$~0.006~(stat.)~$\pm$~0.006~(sys.)~MeV and
    0.501~$\pm$~0.003~(stat.)~$\pm$~0.006~(sys.)~MeV
    for the detectors A and D, respectively
    (see Methods ``Instrumental calibration'').
The hypothesis that the line originates from the background is thus rejected,
    of which the line centre energies should be either
    0.583~MeV ($^{208}$Tl) or 0.609~MeV ($^{214}$Bi).   
The continuum is well explained
    by the combination of photo-absorption
    and Compton scattering of $\sim$~0.51~MeV photons,
    and supports the interpretation of the annihilation line from positrons (Fig.\ref{fig:fig3}).  

The time profile of the annihilation signal (Fig.~\ref{fig:fig3})
    showed an exponentially decaying component
    with time constants of 4--10~s in both the detectors.
A following delayed component was also detected albeit only in the detector A,
    of which the profile is fitted by a Gaussian
    with a peak at $t_\mathrm{peak} =$~34.5~$\pm$~1.0~s
    and a width of 13.2~$\pm$~1.0~s (1$\sigma$).

What produced these positrons and how?
A potential scenario is that
    electron-positron pairs are produced by high-energy gamma rays
    in the electron acceleration process~\cite{GUREVICH1992463, 2012JGRA..117.2308D}.
However, the lack of detection of
    high-energy seed gamma rays ($>$~3~MeV, Fig.~\ref{fig:fig3}a, b) rejects the scenario.
In addition, the environmental electric field
    measured on the ground 
    was upwards during the annihilation signal (below $\sim -3$~kV~m$^{-1}$),
    and hence positrons should not be accumulated towards the ground,
    and the annihilation line should not have been enhanced. 
Consequently, the most natural interpretation of the present data
    is the photonuclear reactions~\cite{2007Ge&Ae..47..664B, 2010JGRA..115.0E19C, JGRA:JGRA20239, 2014PhRvD..89i3010B}; \textit{i.e.},
    a burst (or flash) of the lightning-triggered gamma-ray photons,
    which caused the initial instrumental undershoot,
    collided with atmospheric nuclei and initiated nuclear reactions. 
The atmospheric photonuclear reactions
    $^{14}$N~($\gamma$,~n)~$^{13}$N and
    $^{16}$O~($\gamma$,~n)~$^{15}$O
    generate fast neutrons with a kinetic energy of $E_0 \sim$~10~MeV
    and unstable radioactive isotopes,
    which generate positrons in beta-plus decays.

Produced fast neutrons undergo moderation and diffusion
    down to the epithermal energy ($\sim$~eV)
    via multiple elastic scatterings with atmospheric molecules,
    particularly nitrogen (see Methods ``Neutron propagation'').
During this process,
    96\% of neutrons disappear via charged-particle productions
    $^{14}$N~(n,~p)~$^{14}$C,
    producing quasi-stable carbon isotopes $^{14}$C
    (a half life of 5,730 years)
    without emitting any strong gamma rays,
    while the other 4\% are radiatively captured by atmospheric nitrogen
    or matters on the ground, including those around detectors.
The nuclei that captured a neutron
    promptly emit multiple de-excitation gamma-ray lines,
    e.g., $^{14}$N~(n,~$\gamma$)~$^{15}$N.
The theoretical capture rate
    exponentially decays with a timescale of 56~ms,
    which is consistent with the decay constants of 40--60~ms observed in the sub-second radiation.
The simulated de-excitation gamma-ray spectra for our detectors,
    in which the atmosphere, surrounding materials,
    and detector energy resolution are taken into account,
    are also found to be consistent with the observed data (Fig.~\ref{fig:fig2}).
Notably the sharp cutoff at $\sim$~10~MeV
    is explained to be caused by the lack of nuclear lines above this energy
    (see Methods ``Neutron capture'').

The other major products $^{13}$N and $^{15}$O
    gradually decay into stable $^{13}$C and $^{15}$N via beta-plus decays,
    $^{13}$N~$\rightarrow\ ^{13}$C~$+$~e$^+ + \nu_\mathrm{e}$
    (a half life of 598~s) and
    $^{15}$O~$\rightarrow\ ^{15}$N~$+$~e$^+ + \nu_\mathrm{e}$
    (122~s), respectively.
A region, or ``cloud'', filled with these isotopes
    emits positrons for more than 10 minutes
    and moves by wind above our detectors
    without experiencing much diffusion due to a low mobility of the isotopes.
A positron emitted from $^{13}$N or $^{15}$O
    travels in the atmosphere for roughly a few metres,
    quickly annihilates in meeting an ambient electron,
    and radiates two 0.511~MeV photons,
    whose atmospheric mean free path is $\sim$~89~m.
The epoch $t_\mathrm{peak}$ detected at the detector A
    is consistent with the wind velocity and direction at the day 
    (see Fig.~\ref{fig:fig1} and Methods ``Ambient wind flow'').
The decaying phase ($t <$~10~s, Fig.~\ref{fig:fig3})
    observed with the detectors A and D
    is interpreted as a consequence of beta-plus decay of unstable radioisotopes,
    which were produced around the detectors by the photonuclear reactions
    at the initial radiation flash
    (e.g., $^{27}$Si in the ground with a half life of 4.15~s).

Here we estimate the densities of positrons and neutrons generated by the lightning
    with a help of Monte Carlo simulations,
    on the basis of the photonuclear reactions and subsequent physical processes.
We find with our simulations that
    the obtained annihilation spectrum is well explained by the cloud
    located $\sim$~80~m away from the detector
    (see Methods ``Positrons and annihilation'').
Since the number of the delayed 0.511~MeV photons obtained by the detector A is
    $N_{511} =$~(1.4~$\pm$~0.2)~$\times$~10$^3$
    during the period $\Delta t =$~52~s for 11~$< t <$~63~s,
    the corresponding time-integrated beta-plus-decay density for the period
    is estimated to be $n_{\beta +} =$~3.1~$\times$~10$^{-3}$~cm$^{-3}$,
    assuming a simplified cylindrical volume $V$ for the cloud
    with a horizontal radius $R_\mathrm{d} =$~220~m (see Methods ``Ambient wind flow")
    and fiducial vertical discharge length
    $L_\mathrm{d} =$~1~km~\cite{rakov2003lightning, JGRD:JGRD52300, JGRD:JGRD51099}.
Using the number densities $n_\mathrm{13N}$ and $n_\mathrm{15O}$
    with decay constants $\lambda_\mathrm{13N}$ and $\lambda_\mathrm{15O}$
    of $^{13}$N and $^{15}$O, respectively,
    the beta-plus decay rate is given by
    $S_{\beta +}(t) = \lambda_\mathrm{13N} n_\mathrm{13N}(t) + \lambda_\mathrm{15O} n_\mathrm{15O}(t)$.
In our estimation (see Methods ``Contribution from oxygen''),
    positrons emitted by $^{15}$O amounts 44\% of those by $^{13}$N
    at $t_\mathrm{peak} =$~35~s.
Combining them with the relation $S_{\beta +} \Delta t = n_{\beta +}$,
    we derive $n_\mathrm{13N} =$~1.6~$\times$~10$^{-2}$~cm$^{-3}$ at $t_\mathrm{peak}$,
    and the initial number density of isotopes of
    $n_0 = n_\mathrm{13N}(0) + n_\mathrm{15O}(0) =$~2.6~$\times$~10$^{-2}$~cm$^{-3}$.
Consequently, the total number of the produced neutrons is
\begin{equation}
    N_\mathrm{n} = n_0 V
    = 4 \times 10^{12} 
    \left(\frac{R_\mathrm{d}}{220\ \mathrm{m}}\right)^2 
    \left(\frac{L_\mathrm{d}}{1.0\ \mathrm{km}}\right),
\end{equation}
    which is within the theoretically predicted range of $N_\mathrm{n} =$~10$^{11-15}$
    from the TGF studies~\cite{JGRA:JGRA20239, 2010JGRA..115.0E19C}.

Since the discovery of TGFs with an energy of as high as 20--30~MeV or more,
    the potential nuclear reactions in lightning
    have been hotly discussed~\cite{2010JGRA..115.0E19C, 2007Ge&Ae..47..664B, 2014PhRvD..89i3010B, 2015JPlPh..81d4705D, JGRA:JGRA20239},
    but have never been observationally confirmed before.
We have presented the first unequivocal observational evidence
    for nuclear reactions triggered by lightning.
In fact, we have once detected a similar event before;
    however, the result was marginal at best
    due to limited observational information available at that time~\cite{Umemoto2016}
    (see Methods ``Comparison with the similar event'').
This finding pioneers a research field of atmospheric nuclear reactions in lightning
    studied by positrons and neutrons from the ground.
More importantly,
    these reactions in lightning provide a previously-unknown natural channel on the Earth
    to generate isotopes of carbon, nitrogen, and oxygen
    ($^{13}$C, $^{14}$C, $^{13}$N, $^{15}$N, and $^{15}$O).
The short-lived isotopes $^{13}$N and $^{15}$O emit positrons over our heads,
    of which humans have never been aware before,
    whereas more stable $^{13}$C, $^{14}$C, and $^{15}$N
    contribute to the natural isotope composition on the Earth,
    albeit for a small fraction.
Humans knew up to now only two natural origins of carbon isotopes on the Earth;
    stable primordial isotopes $^{13}$C from geological time,
    originating from the stellar nucleosynthesis~\cite{1995ApJS..101..181W},
    and semi-stable $^{14}$C
    produced via atmospheric interactions with cosmic rays.
Our discovery adds nuclear reactions in lightning as another origin for them.



\begin{figure}
\begin{center}
\includegraphics[width=0.7\hsize,bb=27 20 816 1161]{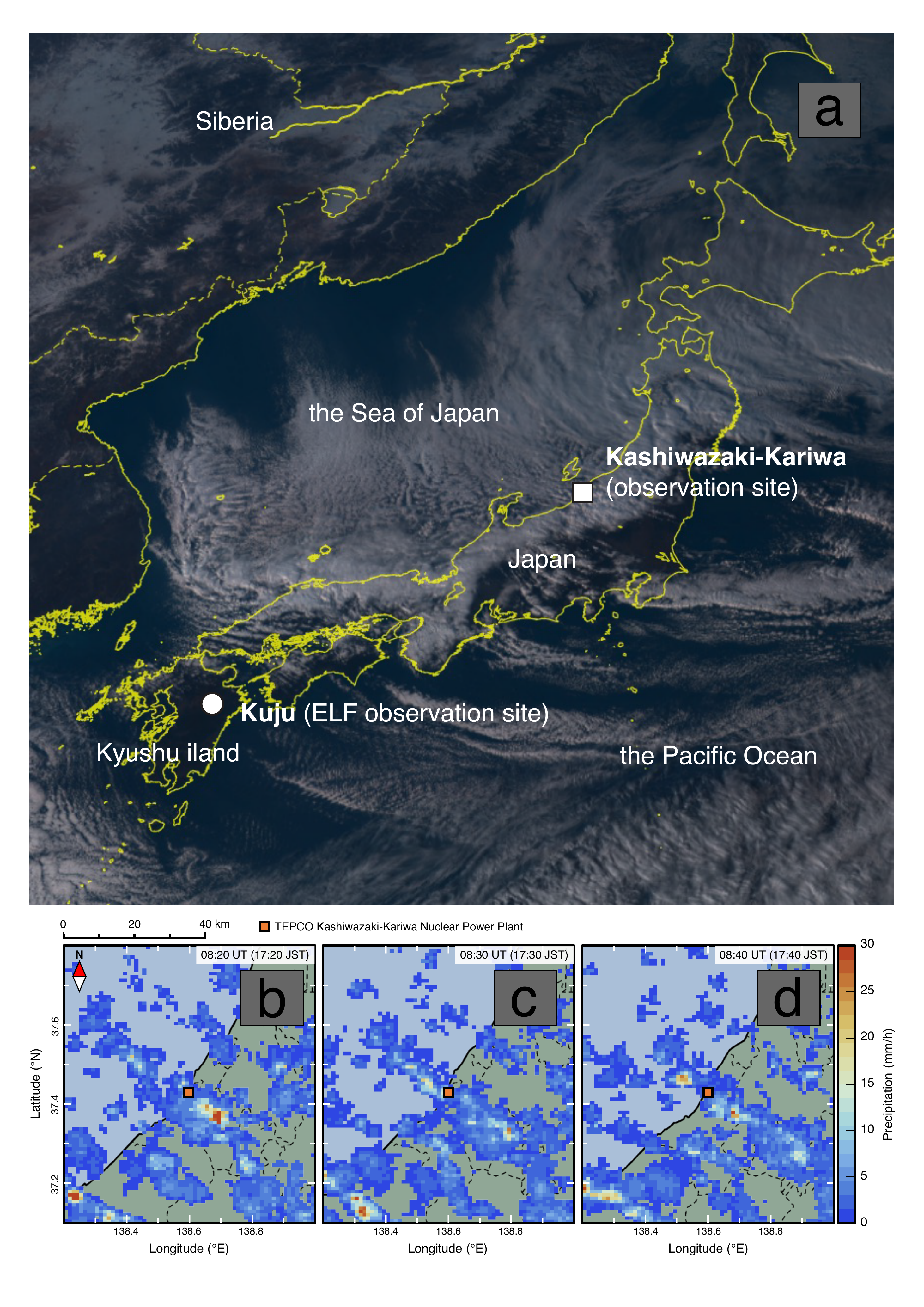}  
\caption{\textbf{Location of the observation sites.}
\textbf{a},
Visible image of the geostationary satellite Himawari 8
    at 06:00 UTC on 6 February 2017.
The square and circle symbols indicate
    Kashiwazaki-Kariwa and Kuju, respectively.
\textbf{b--d},
Precipitation intensity map between 08:20--08:40 UTC on the same day,
    retrieved from the radar system of Japan Meteorological Agency.
Orange square indicates Kashiwazaki-Kariwa Nuclear Power Station.
\label{fig:mfig_himawari}
}
\end{center}
\end{figure}

\begin{figure}
\begin{center}
\includegraphics[width=0.6\hsize,bb=32 15 559 749]{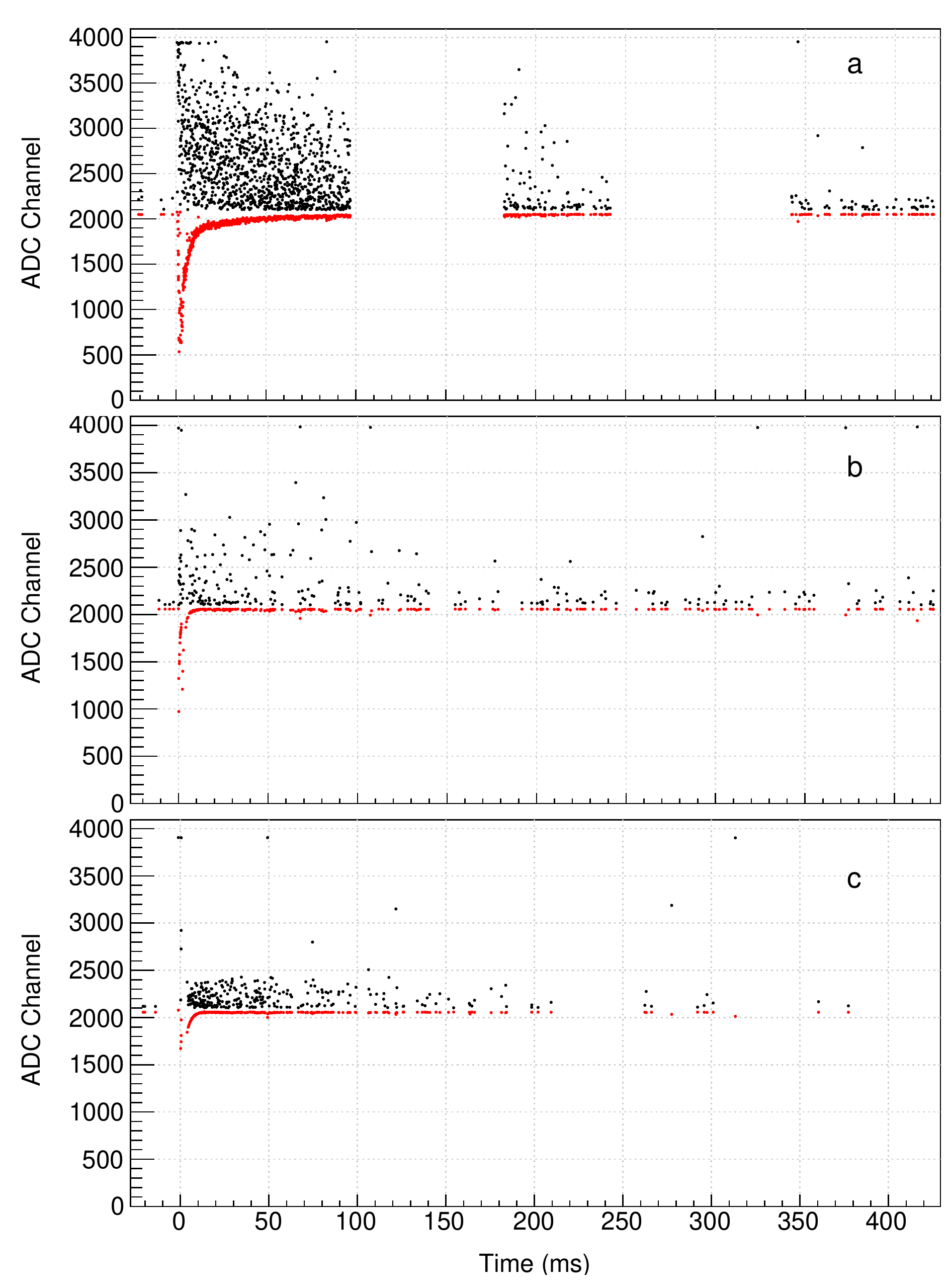}  
\caption{\textbf{Detector response to the initial radiation flash.}
Panels \textbf{a--c} show the time histories of
    the (black dots) maximum and (red) minimum ADC values
    in the ADC-sampled waveforms of the detected photons
    with the detectors A, B, and C, respectively.
Normally the minimum value is equal to the baseline
    ($\sim$~0~V, at ADC~$=$~2050~ch),
    but undershoot was observed in our experiments (see Methods).
The energy 10~MeV corresponds to ADC increases of
    1395, 1218, and 404~ch for the detectors A, B, and C, respectively.
\label{fig:mfig_baseline}    
}
\end{center}
\end{figure}

\begin{methods}



\subsection{GROWTH collaboration.}
The Gamma-Ray Observation of Winter Thunderclouds (GROWTH) project
    is a collaboration to study the high-energy radiation
    from lightning and thunderstorms,
    starting in 2006~\cite{2007PhRvL..99p5002T, 2011JGRD..116.9113T}.
The site of our experiment, Kashiwazaki-Kariwa nuclear power station,
    is located at (37.4267$^\circ$~N, 138.6014$^\circ$~E) in latitude and longitude
    and at an altitude of $\sim$~30--40~m.
We have detected 14 unequivocal events of minute-long gamma-ray enhancements
    from thunderclouds as of 2017 August~\cite{2007PhRvL..99p5002T, 2011JGRD..116.9113T}.
The detectors operated during the winter in 2016--2017
    are summarised in Table~1.
Among them, the detectors A--C were newly deployed in 2016;
    they use a 25~$\times$~8~$\times$~2.5~cm$^3$
    BGO (Bi$_4$Ge$_3$O$_{12}$) scintillation crystal,
    with two photomultiplier tubes (PMTs) HAMAMATSU R1924A attached to for each.
Signals are read out via analogue circuits
    connected to a new ADC board developed with a crowdfunding support.
The detector D has been operated since 2010,
    and consists of a 7.62~cm~$\times \ \phi$~7.62~cm NaI scintillation crystal
    with a PMT R6231 attached to.
Signals of the detector D are read out via another analogue circuit.
In each detector, individual radiation events are recorded
    with time and pulse height.
The effective areas are determined to be 149 and 28~cm$^2$ at 0.511~MeV
    for the detectors A--B and D, respectively,
    with the Monte Carlo simulation using Geant4~\cite{Agostinelli2003},
    in which irradiated photons are monochromatic at 0.511~MeV.
The environmental electric-field was monitored at the position of the detector D.

\subsection{Lightning discharges.}
On 6 February 2017, 
    the Japan lightning-detection network,
    operated by Franklin Japan Co.\ Ltd,
    recorded a negative discharge,
    in which the positive ground charge and negative one in the cloud cancelled each other,
    at 08:34:06.002716165 UTC 
    at (37.432$^\circ$~N, 138.590$^\circ$~E)
    with a peak current of $-33$~kA.
The subsequent positive discharge occurred 23.7~$\mu$s later
    at (37.428$^\circ$~N, 138.588$^\circ$~E)
    with a $+44$~kA peak current.

An associated electromagnetic signal
    was confirmed in the frequency range of 1--100~Hz (ELF range)
    at the Kuju station (33.059$^\circ$~N, 131.233$^\circ$~E) in Japan,
    which is located $\sim$~830~km southwest from the position of the lightning
    (Fig.~\ref{fig:mfig_himawari}).
This ELF observation supports that
    the two bipolar cloud-to-ground discharges occurred simultaneously with this event.

\subsection{Initial flash.}
All our detectors A--D detected a very strong bursting event
    almost coincidentally with the lightning (Fig.~\ref{fig:fig1}).
The detectors A--C digitise the waveform
    of an analogue pulse of the PMT output for 20~$\mu$s
    once the pulse height exceeds the trigger threshold,
    and record the highest and the lowest values.
The former is used to measure the energy of the pulse signal,
    while the latter can be used to monitor the analogue baseline voltage.
In contrast, the detector D has a different analogue circuit from the others
    and does not provide direct information of the baseline voltage,
    which is crucial in this case as discussed below.
For that reason, we exclude the data with the detector D
    for the initial 1 second of the event in our analysis.

At the initial stage of the recorded event,
    we noticed that their baseline voltages were significantly negative.
The level of the baseline gradually settled down to the nominal value
    in 20--40~ms (Fig.~\ref{fig:mfig_baseline}).
This is abnormal, and can not be some technical glitch in the system,
    given that the detectors are completely separate,
    located hundreds of metres apart from one another.

The most plausible interpretation is that
    the detectors received an extremely strong signal
    with the intensity beyond the maximum that they are able to measure,
    lasting for much shorter than 1~ms,
    and then caused the peculiar analogue undershoot lasting for $\sim$~10~ms.
At laboratory we made follow-up experiments using the detectors A--C
    to simulate what these detectors should have experienced at the lightning;
    we applied a bias voltage of 1100~V
    which is higher than the one set during the field observation ($\sim$~900~V)
    to raise the PMT gain by a factor of $\sim$~5.
In this configuration, a cosmic-ray muon-penetration signals
    equivalent to 30--50~MeV energy deposit
    is amplified to 
    150--250~MeV gamma-ray-equivalent charge-output.
We observed the ``peculiar'' undershoot
    with an intensity up to $-1000$ ADC channel
    and a recovery time constant ($\sim$~1~ms),
    both of which are very similar to
    what were observed in the lightning event by the detectors B and C.
This result confirms the interpretation;
    there must have been,
    though not directly measured with our detectors,
    an initial strong radiation flash at the lightning.
In fact, the detector A detected $\sim$~10 events with energies above 10~MeV
    at the very beginning ($<$~2~ms) of the flash,
    before the onset of the undershoot.
The detection further reinforces this interpretation.

In the detector A, the undershoot lasted longer than in the other two.
The level of the undershoot at $t =$~40~ms
    is equivalent to an energy shift of $\sim -0.5$~MeV,
    which changes the energy scale by $>$~50\% below 1~MeV,
    but only by $\sim$~5\% at 10~MeV.
To minimise the effect of this energy-scale shift,
    we extracted and plotted spectra of the sub-second decaying radiation above 1 MeV
    in Fig.~\ref{fig:fig2}.

\subsection{Radiation monitors.}
Radiation monitors are installed in the site
    at $\sim$~300--400~m intervals (Fig.~\ref{fig:fig1})
    and are operated at all times.
Each radiation monitor has
    a spherical ion chamber (IC) filled with $\sim$~14~L argon gas,
    and covers an energy range above 3~MeV with a coarse time resolution of 30~s.

\subsection{Instrumental calibration.}
Energy calibrations are performed
    by a linear fitting of the persistent environmental background emission-lines
    (e.g., Fig.~\ref{fig:fig2})
    of $^{40}$K (1.461~MeV) and $^{208}$Tl (2.615~MeV).
We perform this calibration procedure for the detectors A--C every 30~min
    in order to correct the sensitive temperature dependency
    of light yields of the BGO scintillation crystals ($\sim$~1\% per degree Celsius),
    whereas we do it daily for the NaI of the detector D.
We evaluate the uncertainty of this energy calibration,
    using another background emission line of $^{214}$Bi at 0.609~MeV,
    which appears with rainfall as a component of radon wash-out radiation.
The measured centre energy fluctuates around 0.609~MeV
    with a standard deviation of 0.007~MeV.
Thus, the energies of individual photons are calibrated
    down to an accuracy of 1.1\% (systematic uncertainty),
    or $\sim$~0.006~MeV for 0.511~MeV, in all the detectors A--D.

The information of the absolute timing of individual photons is usually assigned
    from the data of the Global Positioning System (GPS) signals
    with a time resolution of 2~$\mu$s,
    which is the typical time scale of signal waveforms in the detectors A--C.
However, the GPS signals were lost during the event on 6 February.
Hence, the time tags were assigned,
    referring to the internal clock time (Unix time) of our detectors, instead.
We verified the absolute timing uncertainties within 1~s for the three detectors A--C,
    comparing the rising epochs of the initial flash measured with these detectors.
The detector D has a 100~$\mu$s time resolution,
    and the absolute time is calibrated within 100~$\mu$s,
    using the successfully received GPS signals during the event.

\subsection{Neutron propagation.}
The physical processes that follow the photonuclear reactions
    are schematically illustrated in Fig.~\ref{fig:mfig_illustration}.
The photonuclear reactions,
    $^{14}$N~($\gamma$,~n)~$^{13}$N and
    $^{16}$O~($\gamma$,~n)~$^{15}$O, 
    kick out fast neutrons of atmospheric nitrogen and oxygen
    with a typical kinetic energy of $\sim$~10~MeV~\cite{2010JGRA..115.0E19C, 2014PhRvD..89i3010B, 2015JPlPh..81d4705D, JGRA:JGRA20239}.
In the following discussion,
    we only consider the most abundant target $^{14}$N,
    given that the cross section of the second most abundant target $^{16}$O
    is relatively small.
The neutron cross section with $^{14}$N, shown in Fig.~\ref{fig:mfig_14N}a,
    has three main processes:
    elastic scattering,
    charged-particle production $^{14}$N~(n,~p)~$^{14}$C,
    and radiative neutron capture $^{14}$N~(n,~$\gamma$)~$^{15}$N.
Incident fast neutrons gradually lose their kinetic energy
    via multiple elastic scatterings,
    whose microscopic cross-section is almost independent of energy
    $\sigma_\mathrm{es} \sim$~10~barns~\cite{international2000handbook, Shibata_JENDL_2011}
    for 0.01--10$^4$~eV,
    and is much larger than that of the other two processes.
As neutrons are moderated and diffused to the epithermal energy ($\sim$~0.1--100~eV),
    the cross sections of charged-particle production and neutron capture
    gradually increase, and neutrons disappear.

The neutron moderation is in a similar situation
    to some nuclear engineering~\cite{IntroductiontoNuclearEngineering3rdEdition}.
Let \textit{lethargy} $\xi$ be the logarithm
    of the inverse ratio of the change of neutron kinetic energy
    in a single elastic scattering,
    from $E_n$ to $E_{n+1}$, where $E_n$ is the energy after $n$ scatterings.
It is approximated as
    $\xi = \ln \left(E_n / E_{n+1} \right) \sim 2 / (A + 2 / 3) =$~0.136,
    where $A =$~14 is the nitrogen mass number.
Since the energy deposit in a single scattering is
    $\Delta E = E_n - E_{n+1} = (1 - \mathrm{e}^{-\xi}) E_n =$~0.127$E_n$,
    the neutron energy decreases from the initial energy $E_0$
    as $E_n =$~0.873$^n E_0$.
At each scattering, a mean free path $\lambda$ is almost independent
    of the neutron energy~\cite{IntroductiontoNuclearEngineering3rdEdition},
\begin{equation}
    \lambda = 23.8\ \mathrm{m}
    \left(\frac{\sigma_\mathrm{es}}{10\ \mathrm{barns}} \right)^{-1}
    \left(\frac{\rho}{10^{-3}\ \mathrm{g\ cm}^{-3}} \right)^{-1}
    \left(\frac{A}{14} \right)^{-1},
\end{equation}
    where $\rho$ is the atmospheric nitrogen density.
Therefore, the duration between the two contiguous scatterings is
    $\Delta t_n = \lambda \sqrt{m_\mathrm{n} / 2E_n}$,
    where $m_\mathrm{n}$ is the neutron rest mass energy ($m_\mathrm{n} =$~940~MeV).
The elapsed time until the $n$-th scattering is
    $t_n = \sum_{n'} \Delta t_{n'}$.
The number of the neutrons $N_n$ after the $n$-th scattering
    decreases with a loss in the mean free path $\lambda$ of
    $\Delta N_n = N_n
    \left\{1 - \exp\left[-(\sigma_\mathrm{np} + \sigma_\mathrm{cap})
    / \sigma_\mathrm{es} \right] \right\}$,
    where $\sigma_\mathrm{np}$ and $\sigma_\mathrm{cap}$
    are the cross sections of
    charged-particle production and neutron capture, respectively.
Numerically solving for $E_n$, $t_n$, and $\Delta N_n$ with $n$ (Fig.~\ref{fig:mfig_14N}b),
    the number of the surviving neutrons at $t$ for 5--200~ms is approximately
    $N(t) = N(0) \exp(-t / \tau_n)$
    with the decay constant of $\tau_n =$~56~ms.
Therefore, the neutron disappearing rate via capture is
\begin{equation}
    \frac{dN_\mathrm{cap}}{dt}
    = \frac{\sigma_\mathrm{cap}}{\sigma_\mathrm{np} + \sigma_\mathrm{cap}} \frac{d}{dt}N(t)
    \propto \exp\left(-\frac{t}{\tau_n}\right).
\end{equation} 
The theoretically-calculated ($\tau_n \sim$~56~ms)
    and observed (40--60~ms) decay constants
    are found to be consistent with each other.
In addition, assuming isotropic random-walk scatterings,
    the diffusion distance $\sigma_\mathrm{d}$ is estimated to be
\begin{equation}
    \sigma_\mathrm{d} \sim \lambda \sqrt{n}
    = 260\ \mathrm{m}
    \left(\frac{\lambda}{23.8\ \mathrm{m}} \right)
    \left(\frac{n}{120} \right)^{1/2}.
\end{equation}
It is in the same order as the distance
    between our detectors and the lightning discharges.

\subsection{Neutron capture.}
Moderated neutrons are captured by nuclei
    in the atmosphere, surrounding materials, or detectors.
Then the nucleus promptly radiates
    several gamma-ray photons at discrete energies
    below $\sim$~10~MeV within nanoseconds.
We simulated the expected gamma-ray spectra
    with Geant4 Monte Carlo simulations.
We implemented the BGO scintillation crystals,
    aluminium plates of supporting jigs,
    a detector box made of Acrylonitrile Butadiene Styrene (ABS) resin and polycarbonate,
    and lead blocks below them.
The detectors were placed
    on a 1-m thick flat concrete ground, which imitates the building,
    and in the atmosphere with a uniform density of 1.2~$\times$~10$^{-3}$~g~cm$^{-3}$,
    composed of nitrogen (75.527\% in weight fraction),
    oxygen (23.145\%), argon (1.283\%), and carbon dioxide (0.045\%).
De-excitation gamma rays with an energy above 0.1~MeV
    and branching ratio $>$~10\% for the strongest line
    were generated isotropically and uniformly from
    atmospheric nitrogen ($^{14}$N),
    surrounding materials ($^{27}$Al, $^{28}$Si, and $^{207}$Pb),
    and the BGO crystal itself ($^{70}$Ge, $^{72}$Ge, $^{74}$Ge, and $^{209}$Bi),
    according to the branching ratios
    from Evaluated Nuclear Structure Data File (ENSDF) database
    (\url{http://www.nndc.bnl.gov/ensdf/}).
We compared the observed spectra of the detectors A--C
    with the simulated ones (Fig.~\ref{fig:mfig_promptGammaRay}),
    and found that the cylindrical source geometry
    with a horizontal radius of $R_\mathrm{d} =$~220~m
    and a vertical length of $L_\mathrm{d} =$~1~km
    can reproduce the spectrum with the detector A.
In contrast, a source $\sim$~300~m away from the detectors
    consisting solely of the nitrogen contribution
    was found to roughly reproduce the spectra with the detectors B and C.
These results indicate that neutrons hit matters surrounding the detector A,
    while only de-excitation gamma rays from the atmospheric nitrogen
    reached the detectors B and C.

\subsection{Ambient wind flow.}
The ambient wind flow at an altitude of 85~m was northwesterly (Fig.~\ref{fig:fig1})
    with a velocity of $v_\mathrm{w} =$~17~m~s$^{-1}$,
    and was constant within $\pm$1~m~s$^{-1}$ during the event,
    according to a weather monitor near the detector D,
    operated by the nuclear power station.
The wind information was also confirmed
    with the weather radar images from Japan Meteorological Agency (Fig.~\ref{fig:mfig_himawari}b--d).
The time profile of the delayed annihilation signal (Fig.~\ref{fig:fig3}c)
    is approximated by a Gaussian
    with a peak time of $t_\mathrm{peak} =$~34.5~$\pm$~1.0~s
    and a duration of $\sigma_t =$~13.2~$\pm$~1.0~s (1$\sigma$).
The drifting distance of the positron-emitting cloud
    during the period of 0--$t_\mathrm{peak}$
    is then calculated to be $v_\mathrm{w} t_\mathrm{peak} \sim$~590~m.
It is comparable with the separation
    between the detector A and the place of the discharges.
The wind direction is also consistent with our interpretation.
A typical horizontal size (radius) of the cloud is estimated from the duration to be
\begin{equation}
    R_\mathrm{d} = v_\mathrm{w} \sigma_t
    \sim 220~\mathrm{m}
    \left(\frac{v_\mathrm{w}}{17~\mathrm{m\ s}^{-1}}\right)
    \left(\frac{\sigma_t}{13.2~\mathrm{s}}\right).
\end{equation}

\subsection{Positrons and annihilation.}
We here examine positron emission from radioactive isotopes
    and 0.511~MeV annihilation line-emission.
Positrons are isotropically emitted
    with continuous energy spectra following the beta-plus formula
    with maximum kinetic energies of 1.19~MeV and 1.72~MeV
    from $^{13}$N and $^{15}$O, respectively.
Roughly 97\% of positrons with an initial kinetic energy of $\sim$~1~MeV
    annihilate with non-relativistic electrons via positronium formation,
    after losing their kinetic energy within a few metres by ionising ambient atoms,
    and subsequently emit two back-to-back 0.511~MeV photons.
The remaining $\sim$~3\% directly annihilate in flight
    (direct annihilation of relativistic positrons~\cite{2011RvMP...83.1001P})
    and emit two photons with energies between
    $\sim m_\mathrm{e}c^2 / 2$ and
    $\sim E + 3m_\mathrm{e}c^2 / 2$,
    where $m_\mathrm{e}c^2$ and $E$
    are the rest mass and kinetic energies of positron, respectively.
These photons from direct annihilation by nitrogen- and oxygen-origin positrons
    make a weak continuum up to $\sim$~2.0~MeV and $\sim$~2.5~MeV, respectively.

In order to examine the expected spectrum from annihilation processes,
    we performed Geant4 Monte Carlo simulations
    using the similar setup of de-excitation gamma-ray simulations.
We assumed a cylindrical positron-emitting cloud
    with various distances to its base.
Positrons were generated isotropically and uniformly inside the source volume,
    with a continuum energy distribution of the beta-plus decay,
    taking into account the proportion of the contributions from $^{13}$N and $^{15}$O.
Fig.~\ref{fig:mfig_anihilation} shows the resultant simulated spectra.
We then compared this simulation
    with the observed delayed annihilation signal (Fig.~\ref{fig:fig4}),
    and found that the model with the cloud-base distance of 80~m
    reproduce the observed data best.
Using this distance, we calculated the conversion factor
    of the number of beta-plus decays in a unit volume
    to the detected annihilation photons at the 0.511~MeV line to be
    $N_{511} / n_{\beta +} =$~4.5~$\times$~10$^5$~cm$^3$,
    assuming the horizontal radius $R_\mathrm{d} =$~220~m
    (see Methods ``Ambient wind flow'').
The total number of beta-plus decay events in a unit volume is calculated to be
    $n_{\beta +} =$~3.1~$\times$~10$^{-3}$~cm$^{-3}$
    from the observed number of delayed annihilation signals
    $N_{511} =$~1.4~$\times$~10$^3$.

\subsection{Contribution from oxygen.}
The two dominant targets of the photonuclear reactions in the atmosphere
    are $^{14}$N and $^{16}$O.
Here we assume the incident gamma-ray spectrum
    to have the same shape as that in the past reported TGFs, which is
    $N(E) \propto E^{-\Gamma} \exp(-E / E_c)$
    for the photon energy $E$, photon index $\Gamma =$~1.4,
    and cutoff energy $E_c =$~6.6~MeV~\cite{PhysRevLett.106.018501}.
Then, the event-number ratio $\eta_\mathrm{prod}$
    of the photonuclear reactions with $^{16}$O to that with $^{14}$N is estimated to be
    $\eta_\mathrm{prod} =$~10.4\%,
    when integrated for the energies of seed gamma rays up to $\sim$~28~MeV,
    using the atmospheric abundances of
    nitrogen (78.08\%) and oxygen (20.94\%),
    and the experimental cross sections
    (e.g., $\sim$~15~mb and $\sim$~10~mb for $^{14}$N and $^{16}$O
    at $\sim$~23~MeV, respectively~\cite{international2000handbook}).
The event rates per unit time of the subsequent beta-plus decays also differ,
    due to a difference of the decay constants between the two isotopes;
    the half-lives of 597.9 and 122.2~sec for $^{13}$N and $^{15}$O
    correspond with the decay constants of
    $\lambda_\mathrm{13N} =$~1.16~$\times$~10$^{-3}$~s$^{-1}$ and
    $\lambda_\mathrm{15O} =$~5.67~$\times$~10$^{-3}$~s$^{-1}$, respectively.
These give the decay rates to be
    $dN_\mathrm{13N}(t) / dt = \lambda_\mathrm{13N} \exp(-\lambda_\mathrm{13N} t)$ and
    $dN_\mathrm{15O}(t) / dt = \lambda_\mathrm{15O} \exp(-\lambda_\mathrm{15O} t)$
    for $^{13}$N and $^{15}$O, respectively.
Therefore, the ratio of the contribution of the annihilation signal of the positrons
    from $^{15}$O to that from $^{13}$N is
\begin{equation}
    \eta (t) = \eta_\mathrm{prod} \frac{dN_\mathrm{15O}(t) / dt}{dN_\mathrm{13N}(t) / dt}
    =\eta_\mathrm{prod} \frac{\lambda_\mathrm{15O}}{\lambda_\mathrm{13N}}
    \exp[-(\lambda_\mathrm{15O} - \lambda_\mathrm{13N})t]
    = 0.51 \exp \left(-\frac{t}{222\ \mathrm{s}}\right).
\end{equation}
It yields $\sim$~44\% at $t_\mathrm{peak} =$~35~s. 

\subsection{Comparison with the similar event.}
An event associated with 0.511~MeV emission,
    similar to the event reported in this paper,
    was once detected previously
    at the same site on 13 January, 2012,
    after a pair of positive and negative discharges~\cite{Umemoto2016}.
At the time of the event,
    only the detector D was operated.
The data acquisition was heavily hampered by the analogue undershoot for $\sim$~200~ms,
    thus studying the sub-second radiation with de-excitation spectra was impossible.
In addition, 
    since the electric-field monitor was not working at that time,
    we were unable to eliminate the pair production scenario entirely.
In contrast, in the event reported in this paper,
    we measured the environmental electric-field at the place of the detector D,
    using a commercial electric field mill BOLTEK EFM-100,
    and found it to be negative during the delayed (annihilation) phase,
    which implies that electrons moved to the ground away from negatively charged clouds,
    and thus generating the 0.511~MeV line
    without emitting 10--20~MeV bremsstrahlung photons is impossible.

\end{methods}

\bibliography{growth_511keV_nat} 

\begin{addendum}
\item We thank the members of the radiation safety group
    of the Kashiwazaki-Kariwa nuclear power station, TEPCO Inc.\ 
    for providing observation sites,
    Dr.\ Hiroko Miyahara, 
    Dr.\ Norita Kawanaka,
    and Dr.\ Hideaki Ohgaki
    for helpful discussions,
    Dr.\ Hiroyoshi Sakurai,
    Dr.\ Megumi Niikura,
    and the Sakurai group members at RIKEN Nishina Center
    for providing BGO scintillation crystals,
    Dr.\ Toru Tamagawa for his project support,
    Mr.\ Gregory Bowers,
    Dr.\ Masashi Kamogawa,
    and Dr.\ David Smith
    for their helpful suggestion to our interpretation, 
    Mr.\ Shigemi Otsuka
    and Mr.\ Hiroshi Kato
    for supporting the detector developments,
    and RIKEN Advanced Center for Computing and Communication
    for use of HOKUSAI GreatWave supercomputing system
    for Monte Carlo simulations.
This research is supported by JSPS/MEXT KAKENHI grant numbers
    15K05115, 
    15H03653, 
    and 16H06006, 
    by SPIRITS 2019 of Kyoto University,
    and by the joint research program
    of the Institute for Cosmic Ray Research (ICRR), the University of Tokyo.
Our project is also supported by crowdfunding named Thundercloud Project,
    using the academic crowdfunding platform ``academist'',
    and we express our sincere gratitude to
    Yutaka Shikano,
    Yasuyuki Araki,
    Makoto T. Hayashi,
    Nobuo Matsumoto,
    Takeaki Enoto,
    Katsuhiro Hayashi,
    Sumitaka Koga,
    Takashi Hamaji,
    Yu-suke Torisawa,
    Sadashi Sawamura,
    Jim Purser,
    Shozo Suehiro,
    Sumio Nakane,
    Masahiro Konishi,
    Hajime Takami,
    and Tomoo Sawara,
    and all the backers of Thundercloud Project.
The data of Himawari 8 in Fig.~\ref{fig:mfig_himawari}a was obtained from
    Science cloud of National Institute of Information and Communications Technology (NICT),
    Data Integration and Analysis System Program (DIAS) by the University of Tokyo,
    Center for Environmental Remote Sensing (CEReS) of Chiba University,
    and Earth Observation Research Center of Japan Aerospace Exploration Agency.
The data of Fig.~\ref{fig:mfig_himawari}b--d was supplied from
    Japan Meteorological Agency 
    and downloaded from the website of Research Institute for Sustainable Humanosphere,
    Kyoto University (\url{http://database.rish.kyoto-u.ac.jp/index-e.html}).

\item[Author Contributions] 
T.E., Y.W., Y.F., K.O., K.N., T.Y., T.N., and T.H.
    are responsible for the detector developments, data analyses, and interpretation;
    T.E. is the project leader and wrote the draft of the manuscript;
    Y.W. made a major contribution to the detector development, installation,
    and in particular, analysis;
    Y.F. led the Monte Carlo simulations using Geant4;
    K.N. led the installation of the instruments at Kashiwazaki-Kariwa in 2016;
    T.Y. led the development of new data acquisition system after 2015; 
    D.U. provided the previous data in 2012;
    M.S., Y.S., K.M., and H.T. contributed to the data interpretation.

\item[Author Information] The authors declare no competing financial interests.
Correspondence and requests for materials should be addressed
    to T.E.~(email: teruaki.enoto@gmail.com).


\end{addendum}


\clearpage 

\begin{figure}
\begin{center}
\includegraphics[width=0.8\hsize,bb=30 32 998 1172]{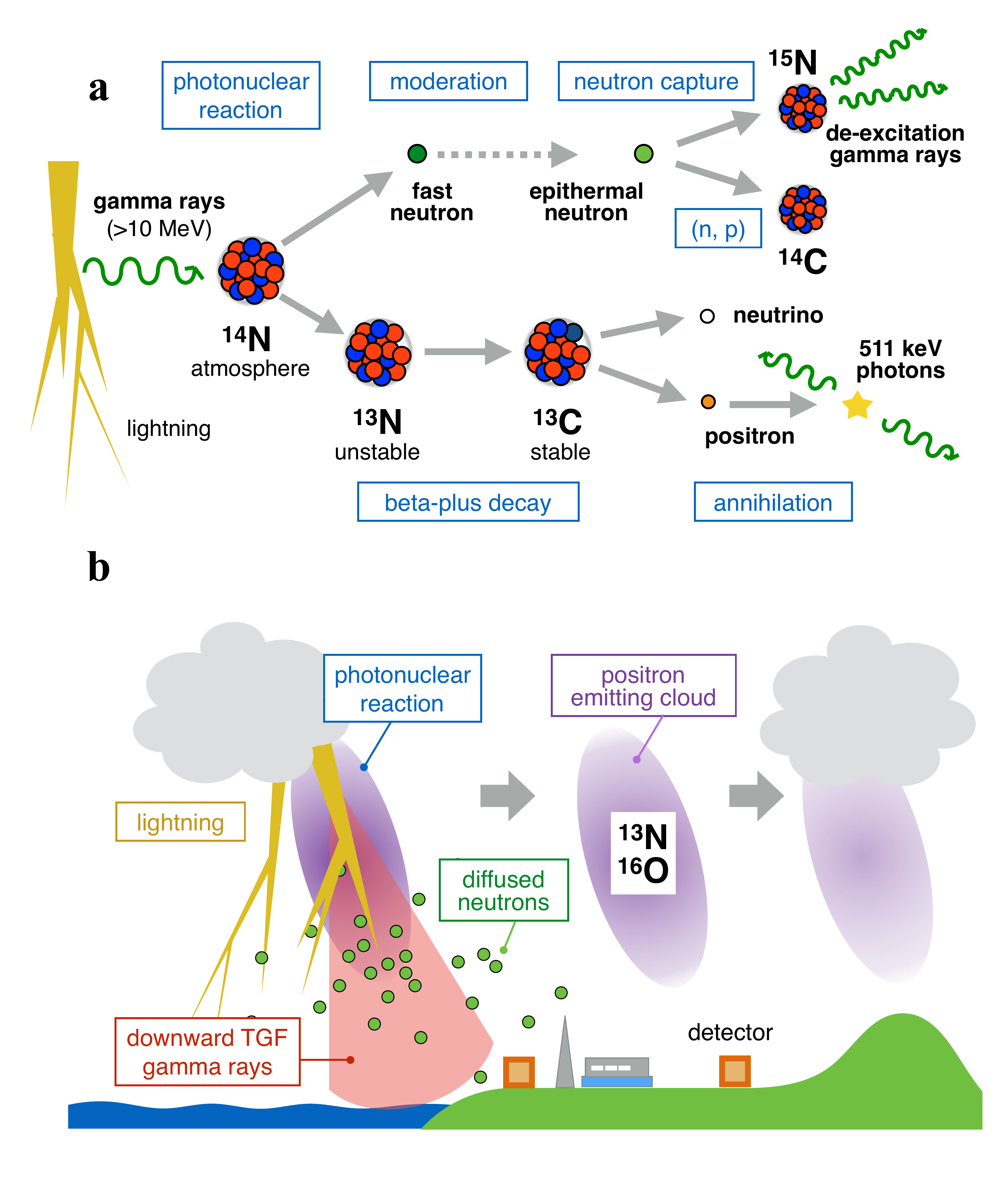}  
\caption{\textbf{Illustration of lightning-triggered physical processes.}
\textbf{a},
Physical processes during a chain of the radiation events
    induced by the photonuclear reactions.
\textbf{b},
Diffusion of neutrons produced in lightning,
    and drift of the positron-emitting cloud.
\label{fig:mfig_illustration}
}
\end{center}
\end{figure}

\begin{figure}
\begin{center}
\includegraphics[width=0.8\hsize,bb=7 0 561 757]{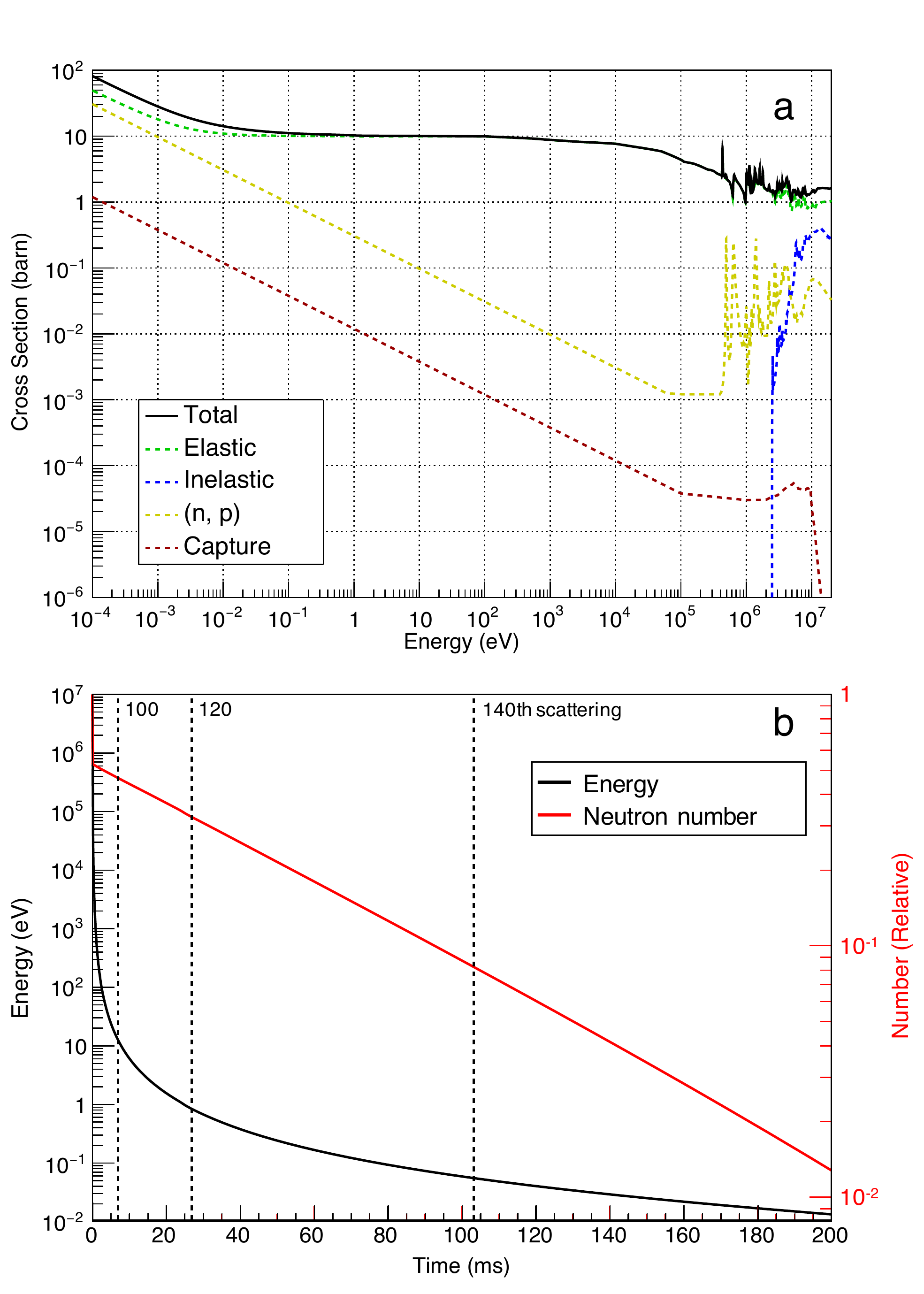}  
\caption{\textbf{Neutron cross-section on nitrogen and time profile of scattered neutrons.}
\textbf{a},
(Black, curved) Neutron cross section on $^{14}$N
    as a function of neutron kinetic energy~\cite{international2000handbook,Shibata_JENDL_2011},
    including both (green) elastic and (blue) inelastic scatterings,
    (yellow) charged-particle production (n,~p),
    and (red) neutron capture.
\textbf{b},
(Black, curved) Kinetic energy and (red) relative number of neutrons
    as a function of time.
The initial energy of neutrons is assumed to be 10~MeV,
    and the initial number of neutrons is normalised to 1.
Dashed lines indicate the times for $n$-th scattering.
\label{fig:mfig_14N}
}
\end{center}
\end{figure}

\begin{figure}
\begin{center}
\includegraphics[width=0.5\hsize,bb=13 0 531 1054]{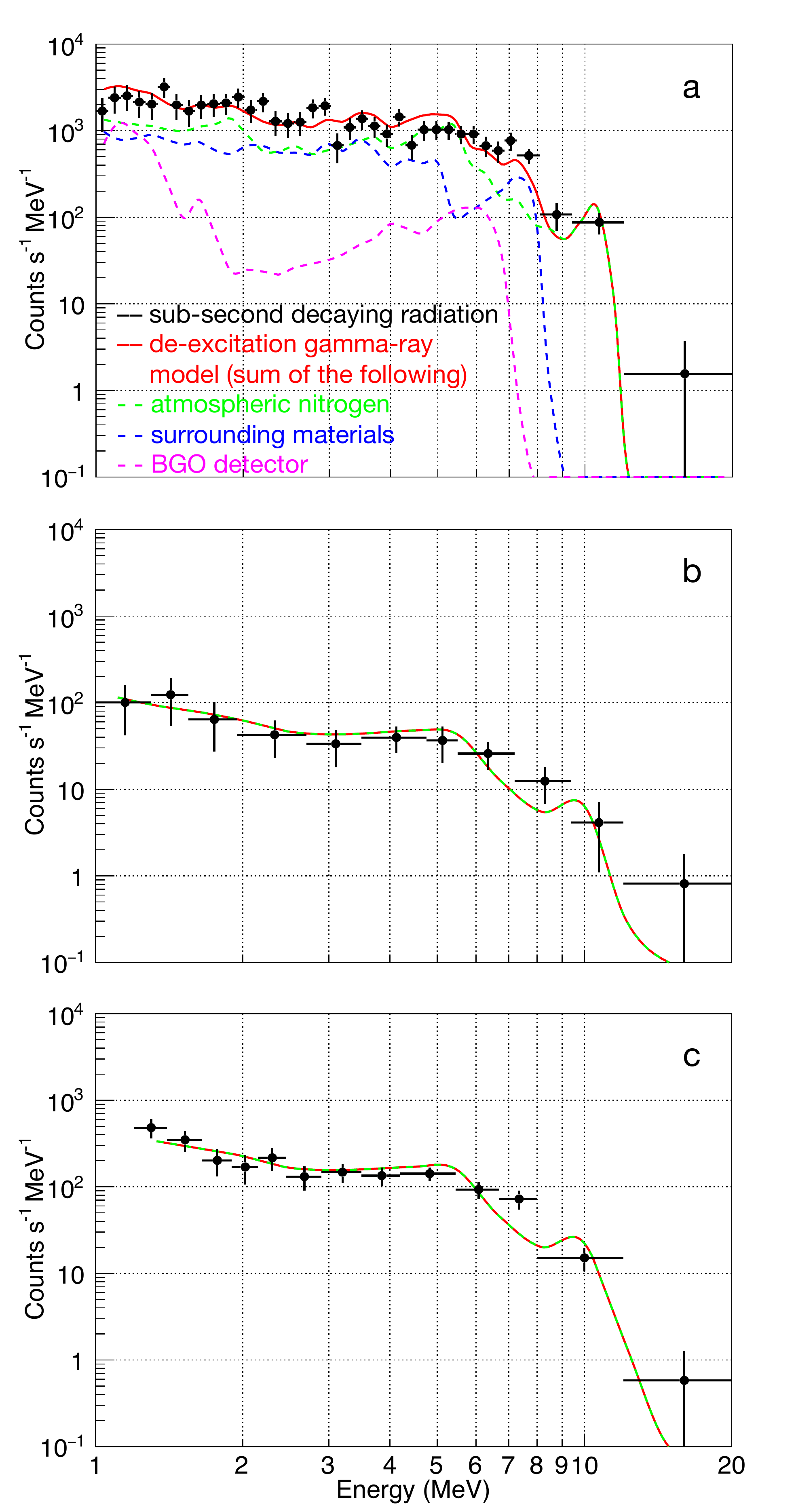}  
\caption{\textbf{De-excitation gamma-ray spectra compared with simulations.}
Panels \textbf{a--c} show the background-subtracted gamma-ray spectra
    of the sub-second radiation
    with black crosses for $\pm 1 \sigma$ errors
    in the detectors A, B, and C, respectively.
The source events are extracted for the periods of
    $t =$~40--100~ms and $t =$~20--200~ms
    for the detectors A and B--C, respectively.
The curves show the Monte Carlo simulations of de-excitation gamma rays from
   (green dashed-line) atmospheric nitrogen,
   (blue dashed) surrounding materials,
   (magenta dashed) detector itself,
   (red solid) and their total.
The simulated spectra are normalised by the total counts above 1~MeV.
\label{fig:mfig_promptGammaRay}    
}
\end{center}
\end{figure}

\begin{figure}
\begin{center}
\includegraphics[width=0.8\hsize,bb=18 0 695 476]{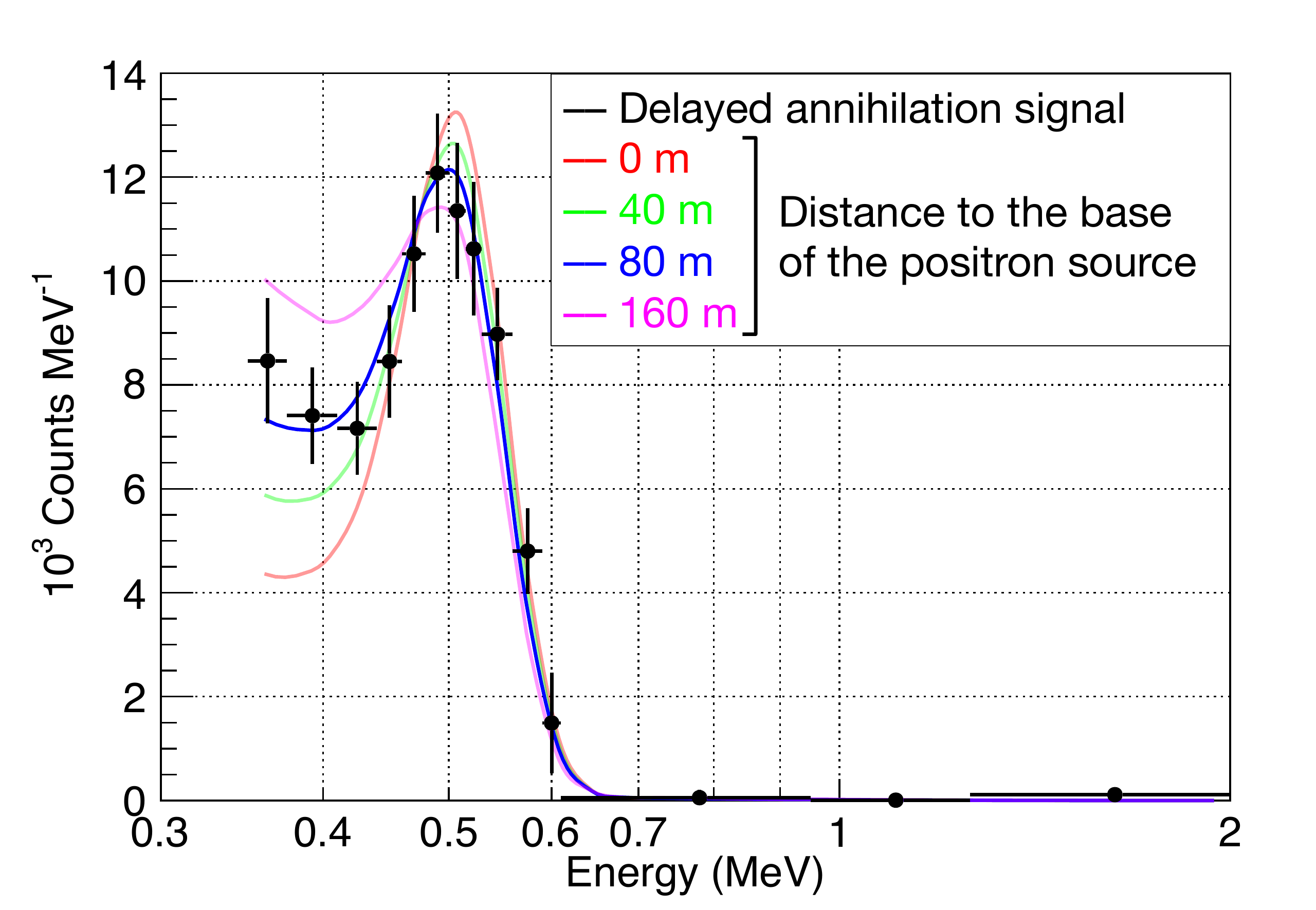}  
\caption{\textbf{Observed annihilation spectrum and simulated models.}
The background-subtracted spectrum in the delayed phase with the detector A,
    accumulated over $t =$~11.1--62.8~s,
    is plotted with black crosses for $\pm 1 \sigma$.
The simulated model curves are overlaid for
    the assumed distances to the base of the positron-emitting cloud of
    (red) 0~m (\textit{i.e.}, the detector is within the cloud),
    (green) 40~m,
    (blue) 80~m,
    and (magenta) 160~m.
The models are normalised by the total counts in the 0.4--0.6~MeV band.
\label{fig:mfig_anihilation}
}
\end{center}
\end{figure}

%

\begin{table}
\centering
\caption{Specifications and obtained values of our detectors
    (The error bars are in $\pm 1 \sigma$).}
\medskip
\small
\begin{tabular}{lcccc}
\hline
Detector & A & B & C & D \\
\hline
Longitude & 138.5960$^\circ$~E & 138.6058$^\circ$~E & 138.6014$^\circ$~E & 138.5907$^\circ~$E \\
Latitude & 37.4211$^\circ~$N & 37.4222$^\circ$~N & 37.4267$^\circ$~N & 37.4200$^\circ$~N \\
Scintillation crystal & Cuboid BGO & Cuboid BGO & Cuboid BGO & Cylindrical NaI \\
Size (cm) & 25~$\times$~8~$\times$~2.5 & 25~$\times$~8~$\times$~2.5 & 25~$\times$~8~$\times$~2.5 & 7.62~$\times \ \phi$~7.62 \\
PMT type & 2~$\times$~R1924A & 2~$\times$~R1924A & 2~$\times$~R1924A & R6231 \\
Energy range (MeV) & 0.35--13.0 & 0.35--13.0 & 1.2--48.0 & 0.2--27.0 \\
\hline
& \multicolumn{4}{c}{Initial radiation flash} \\
Saturated dead time (ms) & $<$~40 & $<$~20 & $<$~20 & $<$~300 \\
\hline
& \multicolumn{4}{c}{Sub-second decaying radiation} \\
Decay constant (ms) & 56~$\pm$~3 & 55~$\pm$~12 & 36~$\pm$~4 & saturated \\
Detected counts & \multirow{2}{*}{1530} & \multirow{2}{*}{132} & \multirow{2}{*}{177} & \multirow{2}{*}{863} \\
 (20--220~ms) & & & & \\
Energy range (MeV) & $>$~0.35 & $>$~0.35 & $>$~1.20 & $>$~0.20 \\
\hline
& \multicolumn{4}{c}{Prolonged annihilation signal (0.511~MeV line)} \\
Photon counts & 1830~$\pm$~240 & $<$~30 (1$\sigma$) & out of range & 366~$\pm$~50 \\
(Delayed comp.) & (1360 $\pm$ 210) & -- & -- & -- \\
Line centre (MeV) & 0.515~$\pm$~0.008 & -- & -- & 0.501~$\pm$~0.007 \\
Effective area (cm$^2$) & \multirow{2}{*}{149.2} & \multirow{2}{*}{149.2} & \multirow{2}{*}{out of range} & \multirow{2}{*}{28.3} \\
at 0.511~MeV & & & & \\
\hline
\end{tabular}
\end{table}

\end{document}

%% file: original_commands.tex
\ifdefined \aj \else
\newcommand\aj{{AJ}}%
\newcommand\actaa{{Acta Astron.}}%
\newcommand\araa{{ARA\&A}}%
\newcommand\apj{{ApJ}}%
\newcommand\apjl{{ApJ}}%
\newcommand\apjs{{ApJS}}%
\newcommand\ao{{Appl.~Opt.}}%
\newcommand\apss{{Ap\&SS}}%
\newcommand\aap{{A\&A}}%
\newcommand\aapr{{A\&A~Rev.}}%
\newcommand\aaps{{A\&AS}}%
\newcommand\azh{{AZh}}%
\newcommand\baas{{BAAS}}%
\newcommand\caa{{Chinese Astron. Astrophys.}}%
\newcommand\cjaa{{Chinese J. Astron. Astrophys.}}%
\newcommand\icarus{{Icarus}}%
\newcommand\jcap{{J. Cosmology Astropart. Phys.}}%
\newcommand\jrasc{{JRASC}}%
\newcommand\memras{{MmRAS}}%
\newcommand\mnras{{MNRAS}}%
\newcommand\na{{New A}}%
\newcommand\nar{{New A Rev.}}%
\newcommand\pra{{Phys.~Rev.~A}}%
\newcommand\prb{{Phys.~Rev.~B}}%
\newcommand\prc{{Phys.~Rev.~C}}%
\newcommand\prd{{Phys.~Rev.~D}}%
\newcommand\pre{{Phys.~Rev.~E}}%
\newcommand\prl{{Phys.~Rev.~Lett.}}%
\newcommand\pasa{{PASA}}%
\newcommand\pasp{{PASP}}%
\newcommand\pasj{{PASJ}}%
\newcommand\qjras{{QJRAS}}%
\newcommand\rmxaa{{Rev. Mexicana Astron. Astrofis.}}%
\newcommand\skytel{{S\&T}}%
\newcommand\solphys{{Sol.~Phys.}}%
\newcommand\sovast{{Soviet~Ast.}}%
\newcommand\ssr{{Space~Sci.~Rev.}}%
\newcommand\zap{{ZAp}}%
\newcommand\nat{{Nature}}%
\newcommand\iaucirc{{IAU~Circ.}}%
\newcommand\aplett{{Astrophys.~Lett.}}%
\newcommand\apspr{{Astrophys.~Space~Phys.~Res.}}%
\newcommand\bain{{Bull.~Astron.~Inst.~Netherlands}}%
\newcommand\fcp{{Fund.~Cosmic~Phys.}}%
\newcommand\gca{{Geochim.~Cosmochim.~Acta}}%
\newcommand\grl{{Geophys.~Res.~Lett.}}%
\newcommand\jcp{{J.~Chem.~Phys.}}%
\newcommand\jgr{{J.~Geophys.~Res.}}%
\newcommand\jqsrt{{J.~Quant.~Spec.~Radiat.~Transf.}}%
\newcommand\memsai{{Mem.~Soc.~Astron.~Italiana}}%
\newcommand\nphysa{{Nucl.~Phys.~A}}%
\newcommand\physrep{{Phys.~Rep.}}%
\newcommand\physscr{{Phys.~Scr}}%
\newcommand\planss{{Planet.~Space~Sci.}}%
\newcommand\procspie{{Proc.~SPIE}}%
\fi